\documentclass[pra,reprint,showpacs,superscriptaddress]{revtex4-1}
\usepackage{graphicx}
\usepackage{bm}
\usepackage{color}
\usepackage[USenglish]{babel}
\usepackage{amsmath}
\usepackage{multirow}
\usepackage{braket}
\usepackage{natbib}
\usepackage{hyperref}
\usepackage{tkz-euclide}
\usepackage[normalem]{ulem}
\usetikzlibrary{calc,patterns,angles,quotes}
\hypersetup{
	colorlinks=true,
	linkcolor=blue,
	citecolor=blue,
	urlcolor=blue
}


\newcommand{\Xstate}{X$^1\Sigma^+$}
\newcommand{\astate}{a$^3\Sigma^+$}

\begin{document}
	
	
	\title{High partial-wave Feshbach resonances in an ultracold $^6$Li-$^{133}$Cs mixture}
	
	\author{Bing Zhu}\email{bzhu@physi.uni-heidelberg.de}
	\affiliation{Physikalisches Institut, Universit\"at Heidelberg, Im Neuenheimer Feld 226, 69120 Heidelberg, Germany}
	\affiliation{Hefei National Laboratory for Physical Sciences at the Microscale and Department of Modern Physics, and CAS Center for Excellence and Synergetic Innovation Center in Quantum Information and Quantum Physics, University of Science and Technology of China, Hefei 230026, China}
	\author{Stephan H\"afner}		
	\author{Binh Tran}
	\author{Manuel Gerken}
	\author{Juris Ulmanis}
	\affiliation{Physikalisches Institut, Universit\"at Heidelberg, Im Neuenheimer Feld 226, 69120 Heidelberg, Germany}
	
	\author{Eberhard Tiemann}
	\email{tiemann@iqo.uni-hannover.de}
	\affiliation{Institut f\"ur Quantenoptik, Leibniz Universit\"at Hannover, Welfengarten 1, 30167 Hannover, Germany}
	
	\author{Matthias Weidem\"uller}
	\email{weidemueller@uni-heidelberg.de}
	\affiliation{Physikalisches Institut, Universit\"at Heidelberg, Im Neuenheimer Feld 226, 69120 Heidelberg, Germany}
	\affiliation{Hefei National Laboratory for Physical Sciences at the Microscale and Department of Modern Physics, and CAS Center for Excellence and Synergetic Innovation Center in Quantum Information and Quantum Physics, University of Science and Technology of China, Hefei 230026, China}

	\date{\today}
	
    \begin{abstract}
	We measure higher partial wave Feshbach resonances in an ultracold mixture of fermionic $^6$Li and bosonic $^{133}$Cs by magnetic field dependent atom-loss spectroscopy. For the $p$-wave Feshbach resonances we observe triplet structures corresponding to different projections of the pair rotation angular momentum onto the external magnetic field axis.
	We attribute the splittings to the spin-spin and spin-rotation couplings by modelling the observation using a full coupled-channel calculation. Comparison with an oversimplified model, estimating the spin-rotation coupling by describing the weakly bound close-channel molecular state with the perturbative multipole expansion, reveals the significant contribution of the molecular wavefunction at short internuclear distances. Our findings highlight the potential of Feshbach resonances in providing precise information on short- and intermediate-range molecular couplings and wavefunctions. The observed $d$-wave  Feshbach resonances allow us to refine the LiCs singlet and triplet ground-state molecular potential curves at large internuclear separations. 
    \end{abstract}
	
	\maketitle
	
	\section{Introduction}
	
	In the field of ultracold quantum gases, tunable interactions via magnetic Feshbach resonances (FRs) are a versatile tool to study of few- and many-body physics, e.g. ultracold molecules~\cite{Quemener2012}, Efimov physics~\cite{Braaten2007}, and the Bardeen-Cooper-Schrieffer (BCS) superfluid (SF) to Bose-Einstein condensate (BEC) crossover regime \cite{Giorgini2008}. While these studies involved pairwise interaction in the $s$-wave channel, interactions at \emph{finite} rotational angular momentum are predicted to lead to intriguing effects such as $p$-wave superfluidity (SF) in liquid $^3$He \cite{Vollhardt2013}, $d$-wave high-$T_C$ superconductivity \cite{Scalapino1995}, and super-Efimov effect in two-dimensional $p$-wave Fermi gases \cite{Naidon2017}. $p$-wave SF may be realized in spin-polarized fermionic atoms interacting via $p$-wave FRs \cite{Ho2005,Ohashi2005,Botelho2005}, where the observed splitting into $m_l=0$ and $m_l=\pm1$ components \cite{Ticknor2004} can give rise to a phase transition between a topologically trivial and non-trivial SF phase \cite{Gurarie2005, Cheng2005}. However, observation of $p$-wave SF in these systems remains elusive due to the short lifetime of $p$-wave Feshbach molecules \cite{Gaebler2007}. Recently, broad $d$-wave resonances, which are promising for realizing $d$-wave SF pairing, have been observed in Refs. \cite{Cui2017, Yao2019}.
	
	For high partial-wave ($l>0$) FRs, anisotropic interactions in the closed channel lift the degeneracy among different projections $m_l$ of $l$ in the lab frame and result in anisotropic scattering processes. This anisotropy can be controlled via tuning the external magnetic fields, which manifests itself as a multiplet structure in the observed FR. One well-known example is the effective spin-spin (\emph{ss}) interaction \cite{Ticknor2004}, arising from both the direct magnetic dipole-dipole interaction (mDDI) and the second-order spin-orbit (\emph{so}) coupling, which splits a $l$-wave FR into $l+1$ resonance positions corresponding to $|m_l|=0,1,\dotsc,l$. The \emph{ss} splitting has been observed in $p$- \cite{Ticknor2004, Pilch2009a, WangP2011, Repp2013, WangF2013, Dong2016} and $d$-wave FRs \cite{Cui2017} and predicted to give rise to a phase transition between polar ($p_x$) and axial ($p_x+ip_y$) SF phases near $p$-wave FRs in spin-polarized degenerate Fermi gases \cite{Gurarie2005,Cheng2005}.
	
	In this work we extend our previous high-precision study of $^6$Li-$^{133}$Cs $s$-wave FRs \cite{Ulmanis2015} to higher partial waves. We observe $p$- and $d$-wave FRs in an ultracold mixture of $^6$Li and $^{133}$Cs via high-precision atom-loss spectroscopy and provide a theoretical description with a full coupled-channel (cc) calculation. 
	
	As we recently demonstrated in Ref. \cite{Zhu2019}, instead of the well-known doublet structures in $p$-wave FRs due to the effective spin-spin (\emph{ss}) interaction \cite{Ticknor2004}, a splitting into triplet structures corresponding to three projections $m_l=-1,0,+1$ of the rotational angular momentum $l=1$ was observed. Such splittings were previously only studied at fairly low fields \cite{Park2012}, where the manifold of the pair rotation $l=1$ is almost degenerate and the non-diagonal terms of the \emph{ss} interaction result in the splitting. The present observation at high fields provides a full control over the angular momentum $l$ and its projection $m_l$ in the scattering process by the external magnetic field. 

	We attribute the observed splitting between $m_l=+1$ and $m_l=-1$ components to the electronic spin-rotation (\emph{sr}) coupling, which is known from molecular spectroscopy of $\Sigma$-states with non-zero electronic spin $S$ \cite{Hund1927,Kramers1929,Vleck1929,LefebvreBrion1986}. This coupling takes the following form in a diatomic $^{2S+1}\Sigma$ molecular state ($S\neq0$)
	\begin{equation}
	H_{sr} = \frac{\gamma}{2 \mu R^2} \mathbf{S} \cdot \mathbf{N} \equiv \gamma\, B(R) \,\mathbf{S} \cdot \mathbf{N}
	\label{eq:srhamiltonian}
	\end{equation}
	(see, e.g., Ref. \cite{Veseth1971}), where $\mu$ stands for the reduced mass of the atomic pair, $R$ the internuclear distance, $B(R)=1/2\mu R^2$ the rotational variable, and $\mathbf{N}$ the total angular momentum excluding the spin, i.e. the rotational angular momentum in $\Sigma$ states. In the following we refer to the pair rotational angular momentum with $N$ and use the atomic units. The dimensionless \emph{sr} coupling constant $\gamma$ contains two main contributions $\gamma=\gamma^{(1)}+\gamma^{(2)}$: The first-order effect $\gamma^{(1)}$ arises from the direct coupling between the electron spins and the magnetic field produced by the rotating charges \cite{Kramers1929} and the second-order effect $\gamma^{(2)}$ accounts for a combined perturbation of the \emph{so} and orbit-rotation interaction \cite{Vleck1929}. Simple estimations of both terms using the fact that the molecular states causing FRs are weakly bound deviate from the experimental observation, indicating significant contributions at short range.
	

Adding the accurate determination of Li-Cs $d$-wave FRs also observed in our experiments gives access to the rotational ladder of $N=0,1,2$ states of the least bound vibrational state and allows us to improve the long-range description of the ground-state molecular potential curves.
	
This paper is structured as follows: In Sec. \ref{sec:prep} we describe the atom-loss Feshbach spectroscopy on the $p$- and $d$-wave FRs. To understand the observed triplet splittings in $p$-wave FRs we estimate the \emph{sr} coupling constant at large internuclear distance using a simple model in Sec. \ref{sec:estimate_ss_sr}. The cc calculation is described and the refined molecular potentials are presented in Sec.~\ref{sec:ccc}.  We conclude the paper in Sec. \ref{sec:conclusion}.
	
	\section{Atom-loss Feshbach spectroscopy}
	\label{sec:prep}

   \begin{figure}[t]
   	\centering
   	\includegraphics[width=0.5\textwidth]{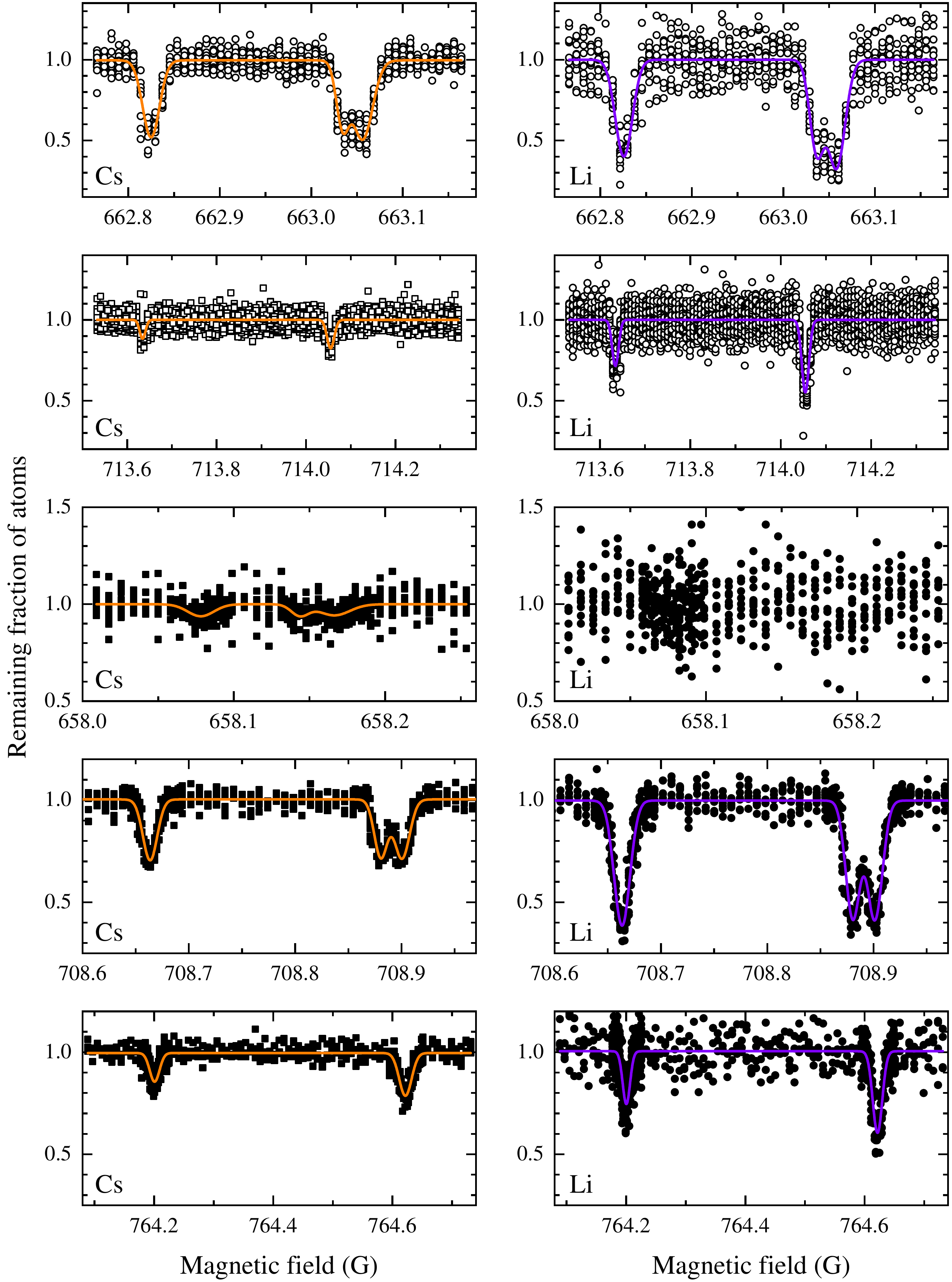}
   	\caption{Li-Cs $p$-wave FRs in the Li$\left|1/2,1/2\right\rangle \oplus$ Cs$\left|3,3\right\rangle$ (open symbols) and Li$\left|1/2,-1/2\right\rangle \oplus$ Cs$\left|3,3\right\rangle$ (filled symbols) channels observed in the remaining fraction of Cs (squares, left column) and Li (circles, right column) atoms. Each row corresponds to one resonance and triplet (doublet) structures are observed in $p$-wave FRs close to 658~G, 663~G, and 709~G (714~G and 764~G) after holding times of 10~s, 0.5~s, and 1~s (5~s and 5~s), respectively. The solid lines are fits of multi-peak Gaussian functions. The fitted resonance positions $B^e_{m_N}$ are listed in Table~\ref{tab:FRs}.
   	}
   	\label{fig:FRLi}
   \end{figure}

	\begin{figure}[t]
		\centering
		\includegraphics[width=0.5\textwidth]{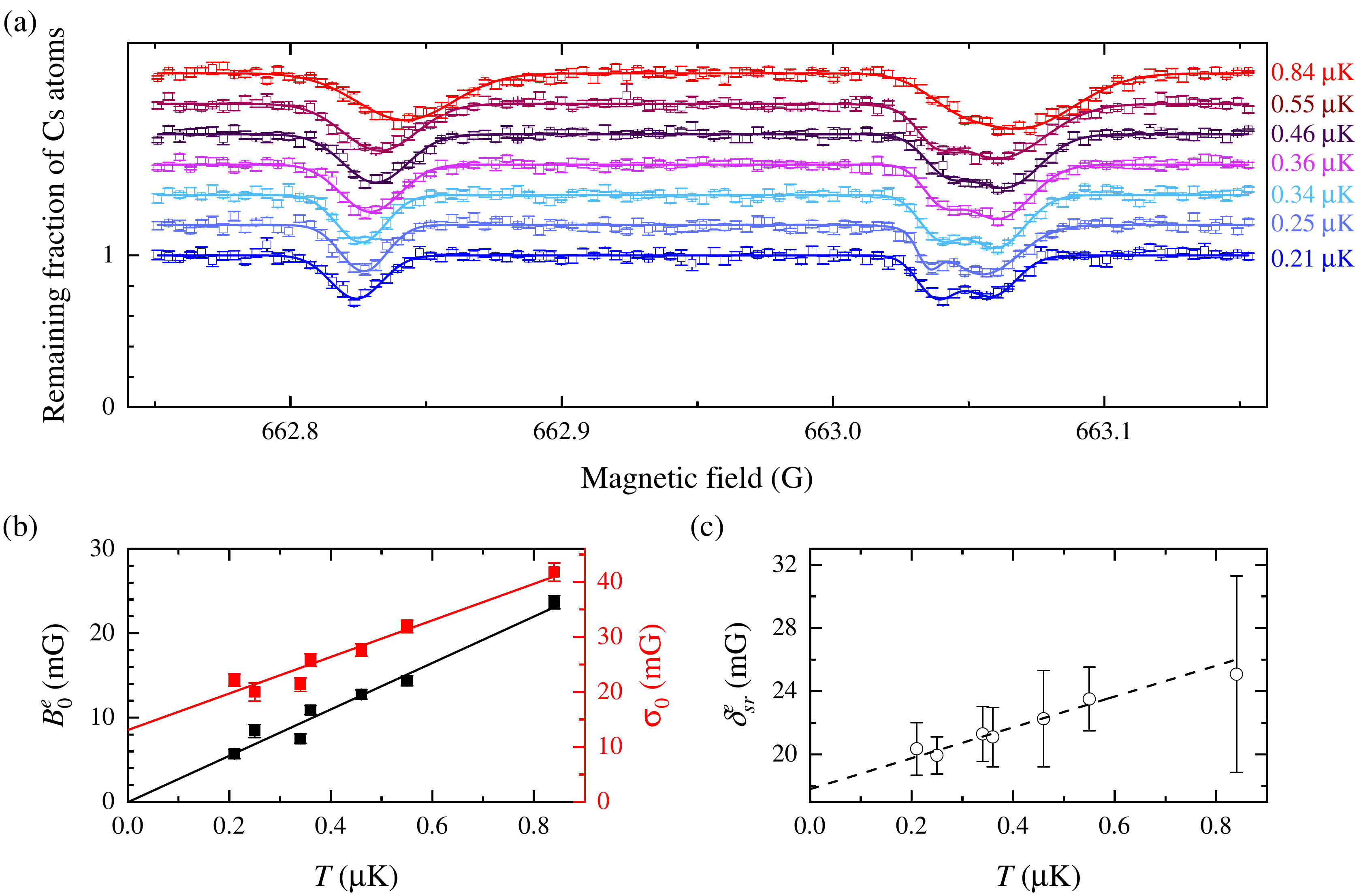}
		\caption{Temperature dependence of the loss features near the 663-G $p$-wave FR. (a) The loss spectra measured at different temperatures. The remaining fraction of Cs atoms after a holding time of 0.2~s is shown as a function of the external magnetic field, where each point is an average of at least 4 experimental runs and the error bars indicate the standard deviations. The spectra are shifted vertically for visibility. Solid curves are fits of triple-Gaussian functions to extract the resonance positions $B^e_{m_N}$ and loss widths $\sigma_{m_N}$. (b) $B^e_{0}$ (black squares) and $\sigma_0$ (red squares) at different temperatures. Error bars come from the fitting in (a). Solid lines are fits of linear functions and $B^e_0$ is referenced to the fitted zero-temperature limit. (c) $\delta^e_{sr}=|B^e_{+1}-B^e_{-1}|$ at different temperatures. Error bars come from the fitting in (a) and the dashed line is a fit of a linear function. (See the main text for detailed discussions.)
		}
		\label{fig:pwaveT}
	\end{figure}
	
	Our experimental procedure for producing an ultracold mixture of Li and Cs is described in Refs. \cite{Pires2014, Ulmanis2015, Zhu2019} . In the end, a mixture of $5\times 10^4$ ($3\times 10^4$) Cs and $3\times 10^4$ ($3\times 10^4$) Li atoms at 430~nK (300~nK) is prepared for the $p$-wave ($d$-wave) measurements. Feshbach spectroscopy is performed by measuring the remaining fraction of Li and Cs atoms after optimized holding times in the trap, ranging from 400~ms (strong $p$-waves) up to 10~s ($d$-wave and one weak $p$-wave resonances), as a function of the magnetic field. We calibrate the magnetic fields by radio-frequency spectroscopy of the nuclear spin-flip transition between the Li $\left|1/2,1/2\right\rangle$ and $\left|1/2,-1/2\right\rangle$ state. The Breit-Rabi formula is used to obtain the magnetic field strength. Its total uncertainty is derived to be 16~mG, caused by long-term drifts, residual field curvature along the long axis of the cigar-shaped trap, and calibration uncertainties.
	
	\subsection{$p$-wave resonances}
	
	We remeasure the five $p$-wave Feshbach resonances (FRs) reported previously in Ref. \cite{Repp2013} with a higher magnetic-field resolution and lower sample temperature and all the loss spectrum after optimal holding times are shown in Fig. \ref{fig:FRLi} with the loss features of both Cs and Li atoms. Triplet (Doublet) structures are observed for the resonances near 658~G, 663~G, and 709~G (714~G and 764~G). Multi-peak Gaussian functions are used to extract out the resonance positions $B_{m_N}^e$ and widths $\sigma_{m_N}$ and the former are listed in Table \ref{tab:FRs} (see also \cite{Zhu2019}). We note that the loss rate close to the 658-G resonance are so small that very shallow loss features are seen only with Cs atoms even after a 10-s holding time.
	
	We have also studied the dependence of the loss features on the cloud temperature near the 663-G resonance by recording the loss spectra at varying trap depths, as shown in Fig. \hyperref[fig:pwaveT]{2(a)}. With increasing temperatures we observe shifts of the resonance positions towards higher magnetic fields and broadening of the loss peaks as expected for $p$-wave FRs \cite{Ticknor2004}. At a temperature of $840nK$ the splitting between $m_N=+1$ and $m_N=-1$ components becomes unresolved as can be seen in Fig. \hyperref[fig:pwaveT]{2(a)}.
	
	In Fig. \hyperref[fig:pwaveT]{2(b)} the resonance positions $B^e_0$ and loss widths $\sigma_0$ are plotted versus the temperature $T$. Similar to Ref. \cite{Gerken2019}, the data is to the lowest order approximated by linear functions, which yield a slope of 28(3)~mG/$\mu$K for the resonance shift and 33(4)~mG/$\mu$K for the loss width with a zero-temperature limit of 13(2)~mG. The temperature-induced resonance shifts for two- and three-body loss processes are about $2.5k_bT/\delta\mu\approx T\times17$~mG/$\mu$K and $k_bT/\delta\mu\approx T\times7$~mG/$\mu$K \cite{Gerken2019}, respectively, with $\delta\mu\approx h\times3$~kHz/mG the differential magnetic moment between the bound molecular and free-atom states for the 663-G resonance. They are both considerably smaller than the fitted slope in Fig. \hyperref[fig:pwaveT]{2(b)}. One possible explanation for this discrepancy is a differential AC stark Shift between the molecular state and the free-atom state coming from the trapping light \cite{Jag2014, Cetina2016a}. The understanding of the loss width is more involved \cite{Gerken2019} and beyond the scope of this work.
	
	Interestingly, in Fig. \hyperref[fig:pwaveT]{2(c)} we observe a temperature-dependent splitting between $m_N=+1$ and $m_N=-1$ components ($\delta^e_{sr}=|B^e_{+1}-B^e_{-1}|$) and a linear fit yields a slope of 10(1)~mG/$\mu$K and a zero-temperature limit of 18(1)~mG. This temperature dependence might indicate that the loss mechanisms for the $m_N=+1$ and $m_N=-1$ components are different, i.e., the difference between the slopes of the temperature-induced shifts of the two- and three-body loss maximums is about $1.5k_bT/\delta\mu\approx10.5$~mG/$\mu$K \cite{Gerken2019}.
	Such an observation implies the importance of using the experimentally measured temperatures when modeling the $p$-wave FRs theoretically.
	
	\subsection{$d$-wave resonances}
	
	\begin{figure}[t]
		\centering
		\includegraphics[width=0.5\textwidth]{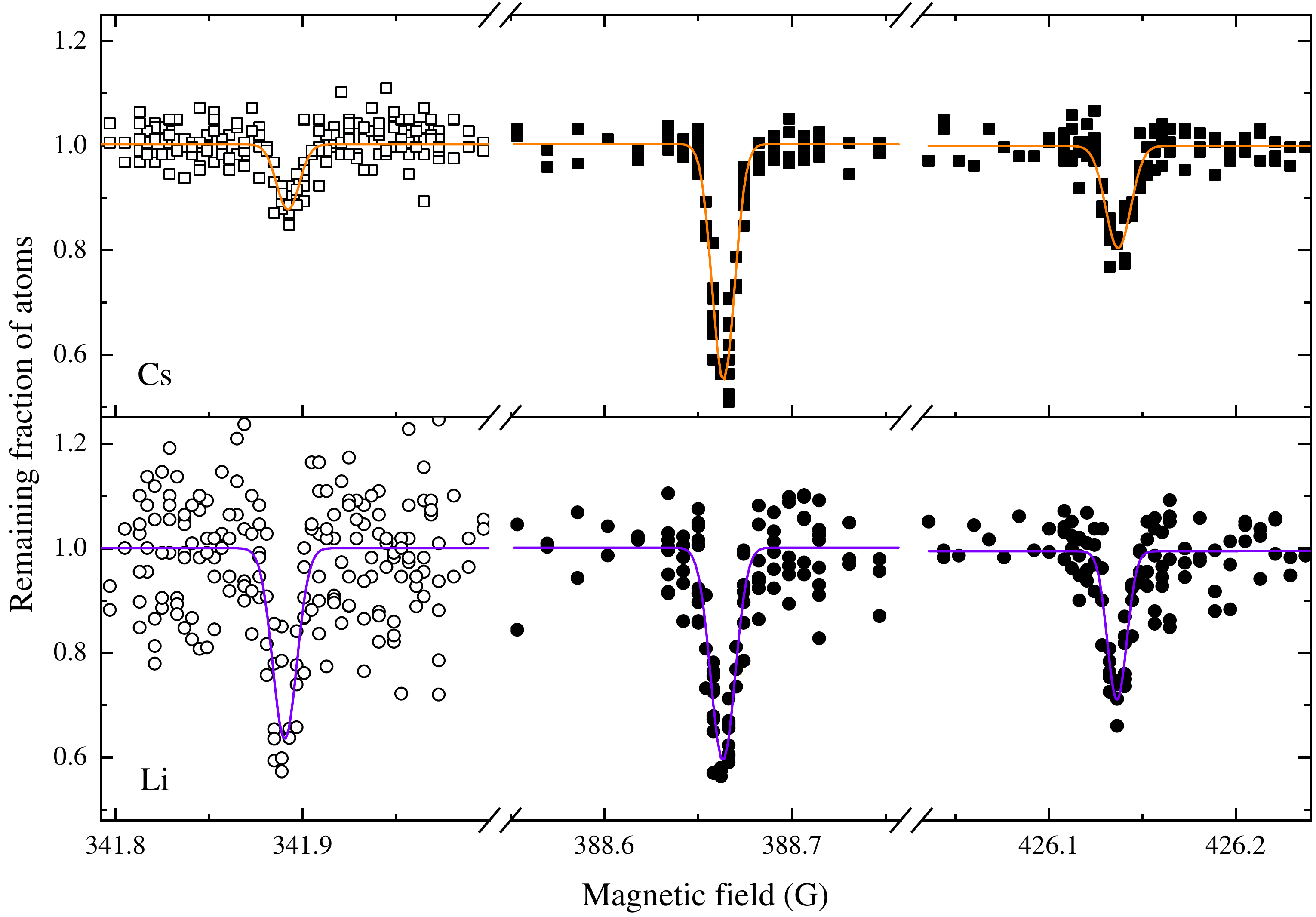}
		\caption{Li-Cs $d$-wave FRs in the Li$\left|1/2,1/2\right\rangle \oplus$ Cs$\left|3,3\right\rangle$ (open symbols) and Li$\left|1/2,-1/2\right\rangle \oplus$ Cs$\left|3,3\right\rangle$ (filled symbols) channels observed in the remaining fraction of Cs (squares) and Li (circles) atoms. Three $d$-wave FRs are observed close to 341.9~G, 388.6~G, and 426.1~G after holding time of 5~s, 5~s, and 8~s, respectively. The magnetic fields are randomly sampled and we reduce the field step size from 8~mG to 2~mG around the loss features. The solid lines are fits of single-peak Gaussian functions.
		}
		\label{fig:dFRCs}
	\end{figure}
	
	To solve a systematic shift between the experimental observations of $p$-wave FR positions and their theoretical description with the molecular potential curves of \cite{Ulmanis2015}, we measure the $d$-wave Li-Cs FRs occurring in the two lowest-energy open channels in the field range of 340~G to 430~G. Instead of the predicted five resonances (see Table \ref{tab:FRs}) we find three of them a couple of hundred mG away from the initial predictions, as shown in Fig. \ref{fig:dFRCs}. No splitting is expected for the observed $d$-wave FRs since the entrance channel is $s$-wave. The loss features are fitted to single-peak Gaussian functions to extract out the resonance positions, which are listed in Table \ref{tab:FRs}. 
    
    In the region around the two unobserved resonances we measure Li-Cs loss rates of $\leq0.02$~$s^{-1}$, which are about a factor of five smaller than those at the resonance positions of the observed ones. This observation agrees with the calculated collision rates around the resonances. However, the atomic loss signal is dominated by the CsCsCs three-body recombination loss rates of approx. 0.15~$s^{-1}$ in this magnetic field range. In addition, close to the predicted Li-Cs $d$-wave FR at 357.92~G we find a Cs-Cs $g$-wave FR~\cite{Berninger2013}, which we determined to be at 358.12(5)~G, leading to enhanced Cs three-body recombination.
    
    \section{A Simple model}
    \label{sec:estimate_ss_sr}
    
     \begin{figure}[t]
    	\centering
    	\includegraphics[width=0.5\textwidth]{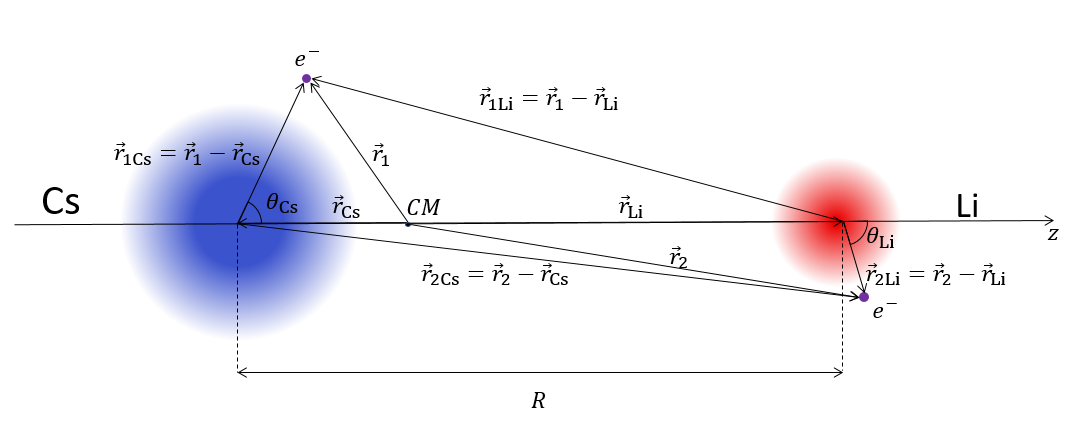}
    	\caption{Schematic drawing for the simple model. The origin of the coordinate system is at the center of mass (\emph{CM}) of the atom pair and the $z$ axis is along the internuclear axis. Each atom is composed of an ionic core at $\vec{r}_\beta = (0,0,z_{\beta})$ and a valence electron at $\vec{r}_i = (x_i,y_i,z_i)$, with $\beta = \mathrm{Li, Cs}$ and $i = 1,2$. The vector from the ionic core $\beta$ to the valence electron $i$ is $\vec{r}_{i\beta} = \vec{r}_i-\vec{r}_\beta=(x_i,y_i,z_i-z_\beta)$. $\theta_\mathrm{Li}$ ($\theta_\mathrm{Cs}$) is defined as the angle from $z$ axis to the vector $\vec{r}_{2\mathrm{Li}}$ ($\vec{r}_{1\mathrm{Cs}}$).
    	}
    	\label{fig:LiCs_sketch}
    \end{figure}
      
   The observed 20~mG splitting between $m_N=+1$ and $m_N=-1$ components of $p$-wave resonances corresponds to an energy splitting about 60~kHz ($B\sim288$~MHz), which is small compared to the \emph{ss} interaction ($\sim600$~kHz). To model this additional splitting arising from spin-rotation coupling, we consider the following simplified model of the Feshbach dimer: As illustrated in Fig. \ref{fig:LiCs_sketch}, each atom is composed of an ionic core and a single valence electron and the pair separated by a distance $R$ is bound via the static Coulomb interactions. To lowest order, the pair wavefunction (WF) can be approximated by the direct product of the atomic WFs of the lithium atom and the cesium atom $\ket{\Phi_\mathrm{2s,6s}}=\ket{\phi^\mathrm{Li}_{2s}}\ket{\phi^\mathrm{Cs}_{6s}}$, both in the electronic ground state with their cores (nucleus and electrons in the inner shells) being separated by a distance $R$. The electronic spins of the two active electrons add to a total spin $S=1$.
    
   
   \subsection{Effective wavefunction}
\begin{figure}[t]
    	\centering
    	\includegraphics[width=0.5\textwidth]{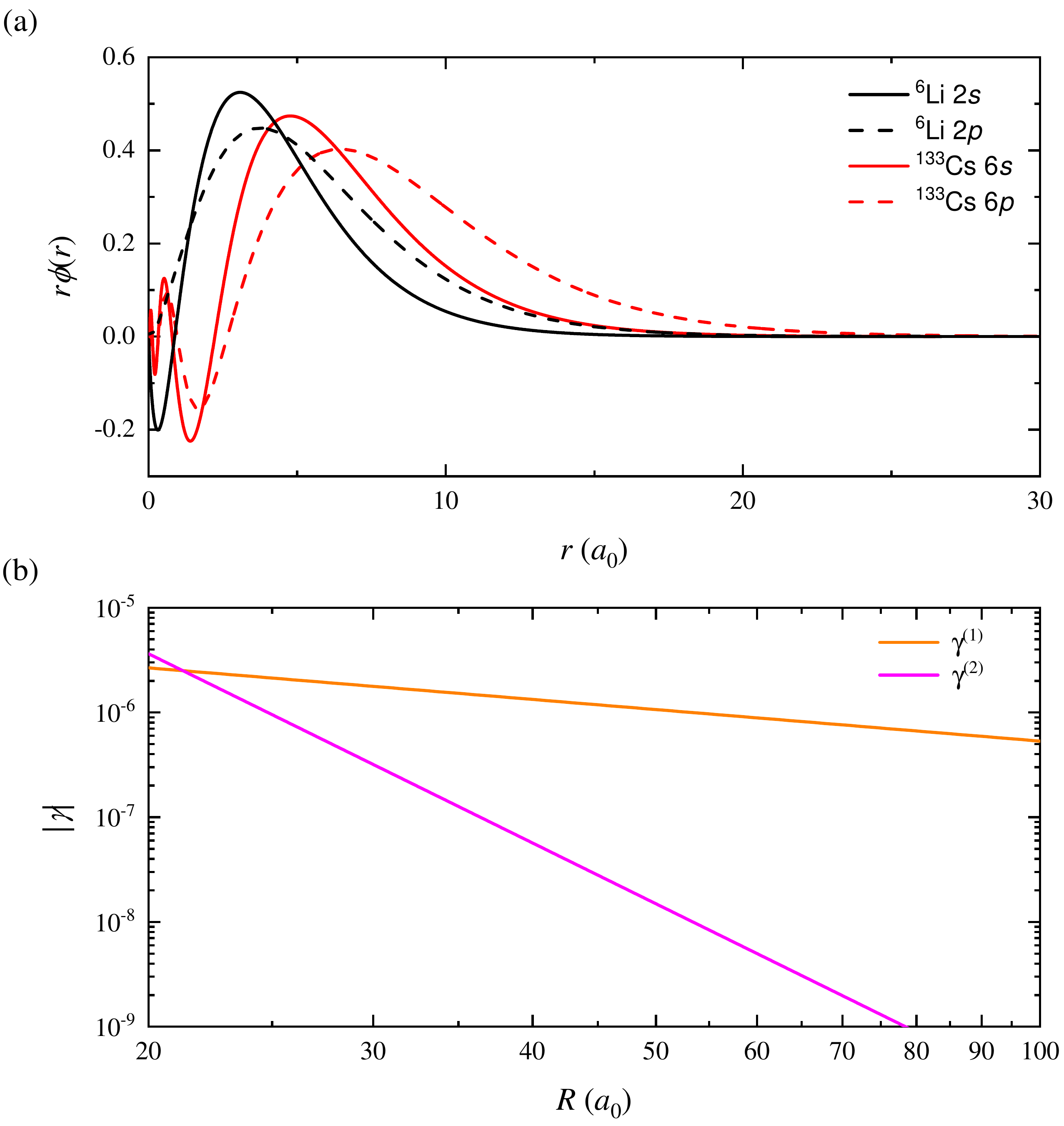}
    	\caption{(a) The unperturbed atomic WFs of the two lowest energy states of $^6$Li and $^{133}$Cs. Here $r$ is the distance between atomic nucleus and valence electron. (b) Log-log plot of estimated first-order and second-order contributions for $\gamma$ at $R>20$~$a_0$. $\gamma^{(1)}$ shows a $R^{-1}$ dependence on $R$ while $\gamma^{(2)}\propto R^{-6}$.}
    	\label{fig:WF_gamma}
    \end{figure}

   At an internuclear distance $R>R_{\mathrm{LR}}\approx20a_0$, the LeRoy radius between the ground state Li and Cs atoms \cite{LeRoy1974}, we approximate the molecular potential between the two atoms by the van-der-Waals interaction $V_\mathrm{vdW}(R)$, resulting from an admixture of the excited state $\ket{\Phi_\mathrm{2p,6p}} = \ket{\phi^\mathrm{Li}_{2p}} \ket{\phi^\mathrm{Cs}_{6p}}$ to the ground-state WF \cite{Marinescu1995, Marinescu1999}. The molecular ground state is then given by 
    \begin{equation}
    	\ket{\Phi; R} = \ket{\Phi_\mathrm{2s,6s}} + \sum C/R^3 \ket{\Phi_\mathrm{2p,6p}}
    	\label{eq:groundstate}
    \end{equation}
    where the sum includes all substates of $\ket{\Phi_\mathrm{2p,6p}}$ and the mixing coefficients $C$ being proportional to the product of the dipole matrix elements between the ground and excited state of lithium and cesium, respectively (see the Appendix \ref{appenA} for more details). The radial WFs can be obtained from the open source library ARC \cite{Sibalic2017} and are depicted in Fig.~\hyperref[fig:WF_gamma]{5(a)} for the two lowest-energy states of $^6$Li and $^{133}$Cs.
    
    \subsection{Spin-rotation coupling}
    Following the derivation as presented in \cite{Brown2003} (see also \cite{Tinkham1955}, where the \emph{ss} and \emph{sr} couplings for the X$^3\Sigma_{g}^{-}$ state of O$_2$ are evaluated), the spin-rotation energy is given by
    Eq. (\ref{eq:srhamiltonian}). It contains two contributions in an effective Hamiltonian model. The first-order term is simply given by the direct coupling of the electron's magnetic moment to the magnetic field associated with the rotation of the charged atomic cores. The \emph{sr} coupling constant then reads
    \begin{equation}
    	\begin{aligned}
    		\gamma^{(1)}(R) = -g \alpha^2 \braket{\Phi;R | \sum_{i,\beta}\frac{\mathbf{S}\cdot\mathbf{s_i}}{S(S+1)}\frac{Z_{i\beta} (z_i-z_\beta) z_\beta}{r_{i\beta}^3} | \Phi;R }.
    	\end{aligned}
    	\label{eq:gamma1}
    \end{equation}
   Here, $g$ is the g-factor of the electron and $\alpha$ is the fine structure constant. The subscripts $i=1,2$ and $\beta = \mathrm{Li,Cs}$ denote the two valence electrons and the atomic cores, respectively. The coordinates are defined in Fig. \ref{fig:LiCs_sketch} and $\mathbf{S} = \mathbf{s}_1 + \mathbf{s}_2$. The effective charges of the atomic cores seen by the valence electrons, including the shielding of the nuclear charge by the inner electrons, is taken as $Z_\mathrm{2Li} = 1.3$, $Z_\mathrm{1Cs} = 6.4$, and $Z_\mathrm{1Li} = Z_\mathrm{2Cs} = 1$ \cite{Clementi1963, Clementi1967}. Due to the large interatomic separation, the contribution of the excited state $\ket{\Phi_\mathrm{2p,6p}}$ to the first-order \emph{sr} energy is negligible within the approximations made here (see Appendix \ref{appenA}).
    
    The second-order contribution to the \emph{sr} energy arises from the finite electronic orbital angular momentum induced by the rotation of the nuclei, which then undergoes spin-orbit coupling involving the electronically excited states admixed to the ground state. The resulting perturbation to the ground-state energy reads \cite{Brown2003}
    \begin{equation}
    	H^{(2)}(R)= \sum \frac{\braket{ \Phi;R | H_\mathrm{so}^e |\Phi_\mathrm{2p,6p}}\braket{ \Phi_\mathrm{2p,6p} |2B(R)\mathbf{L}\cdot\mathbf{N}  | \Phi;R }}{E_\mathrm{Li,2p}+E_\mathrm{Cs,6p}},
    	\label{eq:H2}
    \end{equation}
   where we approximate the excited molecular state with the atomic one $\ket{\Phi_\mathrm{2p,6p}}$ and the sum has to be taken over all its substates. $E_{\mathrm{Li,2p}}$ ($E_{\mathrm{Cs,6p}}$) is the energy of the lithium $2p$ state (cesium $6p$ state) referenced to its atomic ground state. The electronic spin-orbit Hamiltonian is given by $H_\mathrm{so}^e = \sum_i A_i(R)\, \mathbf{l}_i \cdot \mathbf{s}_i$ with $\mathbf{l}_i$ ($\mathbf{s}_i$) the orbital (spin) angular momentum of the electron $i$. The total electronic orbital angular momentum is the sum of individual ones as $\mathbf{L}=\sum_i\mathbf{l}_i$. From Eq. \eqref{eq:H2} we derive the following expression (see Appendix \ref{appenA}).
   \begin{equation}
   \gamma^{(2)}(R)= \frac{1}{2} \sum \frac{\braket{ \Phi;R |A L_- |\Phi_\mathrm{2p,6p}}\braket{ \Phi_\mathrm{2p,6p} |L_+  | \Phi;R }}{E_\mathrm{Li,2p}+E_\mathrm{Cs,6p}}.
   \label{eq:gamma2}
   \end{equation}
   Here $A=\sum_iA_i(R)/2$ and the spin-orbit coupling constants for the two unpaired electrons $A_i(R)$ are approximated by their atomic values $A_\mathrm{Li,2p}$ and $A_\mathrm{Cs,6p}$.

    In Fig. \hyperref[fig:WF_gamma]{5(b)} we show $\gamma^{(1)}$ and $\gamma^{(2)}$ calculated from Eqs. \eqref{eq:gamma1} and \eqref{eq:gamma2}, respectively, as a function of the internuclear distance $R$. Details can be found in the Appendix \ref{appenA}. Both the first-order and second-order effects contribute at most on the order of $5\times10^{-6}$ at $R>20$~$a_0$, which gives rise to an energy on the order of 1~kHz, much smaller than the observed energy splitting of 60~kHz. The deviation of the simple model presented here from the experimental observation mainly comes from the second-order effect, which couples much lower-lying molecular states strongly at short internuclear distances and thus dominates. Short-range contributions need to be taken into account to reproduce the observed triplet structure. In the following we will present a full cc calculation to model the FRs in Li-Cs and the results are compared with our experimental observations.

	\section{Coupled-channel calculation}
	\label{sec:ccc}
	
	\begin{table*}[t]
		\caption{Experimental positions $B^e$ of the Li-Cs $s$-, $p$-, and $d$-wave FRs and comparison to our current cc calculation $\delta=B^e-B^t$. The kinetic energy used in the cc calculation is specified by a temperature $T$, which corresponds to the measured temperatures for the loss-spectroscopy measurements. We assign the good quantum numbers in the closed channel \cite{Repp2013} to all the resonances: the project $m_F$ of the total angular momentum $\mathbf{F}=\mathbf{f}+\mathbf{l}$, $G=S+i_{\text{Cs}}$, and the projection $m_N$ of the pair-rotation angular momentum $N$.}
		\label{tab:FRs}
		\begin{ruledtabular}
			\begin{tabular}{lcccccccc}
				Entrance channel & $B^e$ (G) & $\delta$ (G)& $T$ ($\mu$K) & $m_F$ & $f$ & $G$ & $N$ & $m_N$ \\\hline
				Li$\left|1/2,+1/2\right\rangle$ & 341.891(19) & 0.008 & 0.3 & 7/2 & 9/2 & 7/2 & 2 & 0\\
				$\oplus$ Cs$\left|3,+3\right\rangle$ & 375.367\footnote{Not observed in the experiment. Instead we give the theoretical resonance position $B^t$.} & - & 0.3 & 7/2 & 7/2 & 7/2 & 2 & 0 \\
				
				& 662.822(16) & 0.022 & 0.43 & 7/2 & 9/2 & 7/2 & 1 &0 \\
				
				& 663.036(16) & 0.016 & 0.43 & 9/2 & 9/2 & 7/2 & 1 &+1 \\
				
				& 663.056(16) & 0.019 & 0.43 & 5/2 & 9/2 & 7/2 & 1 &-1 \\
				
				& 713.632(16) & -0.004 & 0.43& 7/2 & 7/2 & 7/2 & 1 &0 \\
				
				& 714.054(16)\footnote{No splitting of $m_N=\pm1$ is resolved in the experiment.} & 0.003 & 0.43 & 9/2 & 7/2 & 7/2 & 1 & +1 \\
				
				& 714.054(16)\footnotemark[1] & -0.006 & 0.43 & 5/2 & 7/2 & 7/2 & 1 & -1 \\
				
				& 842.845(5)\footnote{Derived from rf-spectroscopy of the binding energy of Feshbach molecules.}&-0.006&0.001 & 7/2 & 9/2 & 7/2 & 0 & 0\\
				
				&892.655(30)&0.001&0.1 & 7/2 & 7/2 & 7/2 & 0 & 0\\
				\hline
				
				Li$\left|1/2,-1/2\right\rangle$ &357.920\footnotemark[1] &-&0.3 & 5/2 & 9/2 & 7/2 & 2 & 0 \\
				
				$\oplus$ Cs$\left|3,+3\right\rangle$ &388.663(17) &0.005&0.3 & 5/2 & 7/2 & 7/2 & 2 & 0\\
				
				&426.137(17) &-0.036&0.3 & 5/2 & 5/2 & 7/2 & 2 & 0\\
				
				&658.080(19) &0.046&0.43 & 5/2 & 9/2 & 7/2 & 1 & 0 \\
				
				&658.143(19) &0.040&0.43 & 7/2 & 9/2 & 7/2 & 1 & +1 \\
				
				&658.167(19) &0.039&0.43 & 3/2 & 9/2 & 7/2 & 1 & -1 \\
				
				&708.663(16) &0.015&0.43 & 5/2 & 7/2 & 7/2 & 1 & 0 \\
				
				&708.881(16) &0.009&0.43 & 7/2 & 7/2 & 7/2 & 1 & +1\\
				
				&708.901(16) &0.012&0.43 & 3/2 & 7/2 & 7/2 & 1 & -1 \\
				
				&764.201(16) &-0.021&0.43 & 5/2 & 5/2 & 7/2 & 1 & 0 \\
				
				&764.622(16)\footnotemark[2] &-0.018&0.43 & 7/2 & 5/2 & 7/2 & 1 & +1 \\
				
				&764.622(16)\footnotemark[2] &-0.028&0.43 & 3/2 & 5/2 & 7/2 & 1 & -1 \\
				
				&816.128(20) &0.014&0.3 & 5/2 & 9/2 & 7/2 & 0 & 0 \\
				
				&888.595(5)\footnotemark[3]&-0.005&0.001 & 5/2 & 7/2 & 7/2 & 0 & 0\\
				
				&943.020(50) & -0.035 &0.4 & 5/2 & 5/2 & 7/2 & 0 & 0\\
				\hline
				
				Li$\left|1/2,+1/2\right\rangle$	&704.49(15)&0.16&2 & 5/2 & 9/2 & 7/2 & 1 & 0\\
				
				$\oplus$ Cs$\left|3,+2\right\rangle$ &704.49(15) &0.08 & 2 & 7/2 & 9/2 & 7/2 & 1 & +1\\
				
				&704.49(15)&0.05&2 & 3/2 & 9/2 & 7/2 & 1 & -1\\
				
				&896.62(95)&0.72&2 & 5/2 & 9/2 & 7/2 & 0 & 0 \\
				\hline
				
				Li$\left|1/2,-1/2\right\rangle$	&750.06(15)&0.15&2 & 3/2 & 7/2 & 7/2 & 1 & 0\\
				
				$\oplus$ Cs$\left|3,+2\right\rangle$ &750.06(15)&0.06&2 & 5/2 & 7/2 & 7/2 & 1 & +1\\
				
				&750.06(15)&0.04&2 & 1/2 & 7/2 & 7/2 & 1 & -1\\
				
				&853.85(11)&0.06&2 & 3/2 & 9/2 & 7/2 & 0 & 0\\
				
				&943.5(1.4)&2.25&2 & 3/2 & 7/2 & 7/2 & 0 & 0\\
			\end{tabular}
		\end{ruledtabular}	
	\end{table*}

	\subsection{Hamiltonian}
	
	The full Hamiltonian for the cc calculation reads
	\begin{equation}
	\begin{aligned}
	H=& \ T + \sum_{S=0,1} V_S P_S \\ & + \sum_{\beta} a_\beta(R) \mathbf{s_\beta} \cdot \mathbf{i_\beta} +\sum_{\beta} \mu_B(g_{s,\beta}\mathbf{s_\beta}+g_{i,\beta}\mathbf{i_\beta}) \cdot \mathbf{B} \\ & +\frac{2}{3} \lambda(R) (3 S_Z^2-S^2)  +\frac{\gamma}{2\mu R^2}  \mathbf{S} \cdot \mathbf{N} \, ,
	\end{aligned}
	\label{eq:Hamiltonian}
	\end{equation}
	where $T=-\nabla^2/(2\mu)$ is the relative kinetic energy with the reduced mass $\mu$ of the vibrational and rotational motion. $V_S$ are the molecular potential curves of the singlet ground state \Xstate~and the triplet one \astate, and $P_S$ their projection operators. $S=0,1$ is the total electron spin of the atom pair. 
	
	The third term in Eq. \eqref{eq:Hamiltonian} is the hyperfine coupling with the electronic $\mathbf{s_\beta}$ and nuclear spins $\mathbf{i_\beta}$, where the summation $\beta$ is performed over the two atoms Li and Cs. The functions $a_\beta(R)$ describe the molecular hyperfine coupling  and approach the atomic hyperfine constants at large internuclear separations $R$ \cite{Strauss2010}. Because of missing data of hyperfine splittings of deeply bound levels we keep $a_\beta$ at the respective atomic values. The Zeeman term contains the coupling of the electron and nuclear spins to the homogeneous external magnetic field $B$ with the electronic (nuclear) g-factors $g_{s,\beta}$ ($g_{i,\beta}$) and the Bohr magneton $\mu_B$. 
	
	The last two terms account for the effective \emph{ss} interaction and the \emph{sr} coupling, respectively. The function $\lambda(R)$ in the \emph{ss} interaction includes both the mDDI and the second-order \emph{so} coupling and reads as
	\begin{equation}
	\lambda(R)=-\frac{3}{4}\alpha^2(\frac{1}{R^3}+a_{\mathrm{so1}}\exp(-b_1R)+a_{\mathrm{so2}}\exp(-b_2R)).
	\label{eq:ssparameter}
	\end{equation}
	Here $\alpha$ is the fine-structure constant and $a_{soi}$ and $b_i$ are free parameters to be determined by fitting the experimental double splittings. The \emph{sr} coupling is used in the same form as in Eq. \eqref{eq:srhamiltonian} with the free parameter $\gamma$ to be fitted. Both terms are discussed and compared in detail in Ref. \cite{Zhu2019}, especially the uniqueness of including the \emph{sr} coupling in Eq. \eqref{eq:Hamiltonian}. 
	
	\subsection{Calculation}
	
	The cc calculation is performed as described in \citet{Repp2013} with the extended Hamiltonian from Eq.~\eqref{eq:Hamiltonian}, using atomic hyperfine constants and g-factors of Li and Cs from \citet{Arimondo1977} and atomic masses from tables by  \citet{Audi2003}. The precise determination of FR positions and scattering lengths relies on the knowledge of accurate LiCs molecular potential curves of the electronic ground states \Xstate~and \astate. The molecular potentials are expanded in a power series of internuclear separation $R$ \cite{Gerdes2008}. The coefficients are fitted to simultaneously reproduce 6498 rovibrational transitions from laser-induced fluorescence Fourier-transform spectroscopy \cite{Staanum2007} as well as all known Li-Cs FRs (see Table \ref{tab:FRs}). From the calculation we extract the theoretical resonance positions $B^t$ as the peak positions of the two-body collision rate at a kinetic energy corresponding to the measured atomic cloud temperature for each resonance. 
	
	\subsection{Comparison to measurements}
	
	In Table~\ref{tab:FRs} we give the experimental positions of all the Li-Cs FRs included in the cc calculation and compare them to the calculated results. FRs in the entrance channel Li$\left|1/2,\pm1/2\right\rangle$ $\oplus$ Cs$\left|3,+2\right\rangle$ are taken from Ref. \cite{Repp2013}. The $s$-wave FRs in the two lowest-energy open channels Li$\left|1/2,\pm1/2\right\rangle$ $\oplus$ Cs$\left|3,+3\right\rangle$ are from Ref. \cite{Ulmanis2015} and the $p$- and $d$-wave FRs are from current work. The numbers in brackets give the total uncertainty of the determination of the resonance positions including both the statistical and systematic uncertainties. The theoretical resonance positions $B^t$ are obtained from the cc calculation performed at a relative kinetic energy $k_B T$, according to the experimentally measured temperature $T$, and are given as deviations $\delta=B^e-B^t$. 
	
	The measurement of $s$-, $p$-, and $d$-wave FRs is equivalent to a high-precision spectroscopy of the rotational ladder of the least bound vibrational level and enables us to refine the long-range part of the molecular ground-state potentials. With the optimized molecular potential curves and inclusion of the \emph{sr} coupling we achieve over all 32 FRs a total weighted RMS error of 16~mG, which is an improvement by a factor of two compared to previous theoretical analysis \cite{Pires2014a}.
	
	All the observed resonances are caused by the least vibrational level below the free-atom absolute ground state ($f_{\text{Li}}=1/2+f_{\text{Cs}}=3$) at zero magnetic field, which is a significant triplet-singlet mixture with $\braket{S}\approx0.7$. At high magnetic fields, both the projection $m_f$ of the total angular momentum $f$ (excluding $N$) and the projection $m_N$ of the pair rotation $N$ are good quantum numbers and hence $m_F=m_f+m_N$. $f$ and $N$ are also good quantum numbers. Due to the strong Cs hyperfine coupling, $G=S+i_{\text{Cs}}$ is good as well.
	
	\subsection{Fitted molecular potential curves}
	
	
	The parametrization of the molecular potentials is described for example in \citet{Gerdes2008}. The potentials are represented in three parts: the repulsive short-range part $U_{\mathrm{S}}(R)$, the intermediate range $U_{\mathrm{I}}(R)$ and the asymptotic long range part $U_{\mathrm{L}}(R)$, which are given by the following expressions:
	\begin{equation}
	U_{\mathrm{S}}(R)=A+\frac{B}{R^q}\hspace{0.6cm}\mathrm{for}\hspace{0.6cm} R<R_s ,
	\end{equation}
	\begin{equation}
	U_{\mathrm{I}}(R)=\sum_{k=0}^n d_k \xi(R)^k \hspace{0.6cm}\mathrm{for}\hspace{0.6cm} R_s\leq R\leq R_l ,
	\end{equation}
	\begin{equation}
	\mathrm{with}\hspace{0.2cm} \xi(R)=\frac{R-R_m}{R+b R_m} , \\
	\end{equation}
	and
	\begin{equation}
	U_{\mathrm{L}}(R)=-\frac{C_6}{R^6}-\frac{C_8}{R^8}-\frac{C_{10}}{R^{10}}-...\pm E_{ex}\hspace{0.2cm}\mathrm{for}\hspace{0.2cm} R<R_l, 
	\end{equation}
	where the exchange energy is given by
	\begin{equation}
	E_{ex}=A_{ex} R^\gamma \exp{(-\beta R)}.
	\end{equation}
	It is negative for the singlet and positive for the triplet potential. The long range parameter C$_i$ with $i>10$ are used for smooth connection at R$_l$. 
	
	For the deep ground state we fitted an adiabatic Born-Oppenheimer correction \cite{Tiemann2009} for the simultaneous description of both isotopes of Li. The correction potential U$_{\mathrm{corr}}$(R) is represented as
	\begin{equation}
	U_{\mathrm{corr}}(R) = \left(1-\frac{M_{\mathrm{ref}}}{M}\right) \cdot U_{\mathrm{ad}}(R)
	\label{eq:7}
	\end{equation}
	with
	\begin{equation}
	U_{\mathrm{ad}} (R)= \left(\frac{2R_m}{R+R_m}\right)^n \sum_{i} v_i \cdot \xi(R)^i\; ,\; i=0,1,2,...   
	\label{eq:8}
	\end{equation} 
	with $M_{\mathrm{ref}}$ being the  mass of the selected reference isotope $^{7}$Li and $n$ being the power of R in the leading term of the long range interactions, i.e. $n=6$. 
	
	The parameters of the refined LiCs singlet and triplet molecular potential curves are listed in the Appendix (see Table~\ref{tab:LiCspot}). A computer code in FORTRAN for calculating the potential functions can be found in the supplement of \cite{Pires2014a}. From the new molecular potentials we calculate the singlet and triplet scattering lengths for both Li isotopologues and list them in Table \ref{tab:length}.	They are slightly different from the former reported values \cite{Pires2014a} which reflects the largely extended data set. Because of missing observation of FRs for $^7$Li-$^{133}$Cs, the uncertainties for the corresponding values extrapolated by mass scaling and Born-Oppenheimer correction are larger in comparison to the uncertainties in $^6$Li-$^{133}$Cs.
	
	The parameters in Eq. \eqref{eq:ssparameter} describing the second-order \emph{so} contribution to \emph{ss} interaction are fitted to be:
	$a_{\mathrm{so1}}=-1.99167 $, $b_1= 0.7$, and $a_{\mathrm{SO2}}= -0.012380$, $b_2=0.28 $. The \emph{sr} parameter $\gamma$ in Eq. \eqref{eq:Hamiltonian} has the value $\vert\gamma\vert=0.000566(50)$.  
	
	\begin{table}[t]
		\caption{Scattering lengths $a$ for states \Xstate~and \astate for the isotoptic combinations of $^6$Li-$^{133}$Cs and $^7$Li-$^{133}$Cs in units of Bohr=0.05292 nm.}  
		\label{tab:length}
		\begin{ruledtabular}
			\begin{tabular}{lrr}
				state & $a$ ($^6$Li-$^{133}$Cs) & $a$ ($^7$Li-$^{133}$Cs) \\\hline
				\Xstate	&30.147(50)&45.28(30)\\
				\astate&-34.97(15)&846.4(200)\\
			\end{tabular}
		\end{ruledtabular}	
	\end{table}
	
    \section{Conclusion}
    \label{sec:conclusion}
    In conclusion, we have experimentally and theoretically studied $p$- and $d$-wave FRs in a Li-Cs mixture. By employing all known spectroscopic data for the states \Xstate and \astate and all measured FRs in our experiment we have derived new ground-state potentials in a least squares fit. Limitations in their usage for predicting molecular levels should be expected because FRs were only observed for the isotope pair $^6$Li-$^{133}$Cs, whereas laser spectroscopy \cite{Staanum2007} was mainly performed on the $^7$Li isotope, and only few data on the state \Xstate~for $^6$Li$^{133}$Cs exist. The extension of the Feshbach data to $^7$Li-$^{133}$Cs and the conventional laser spectroscopy covering the isotopologue $^6$Li$^{133}$Cs would be very important to improve the knowledge of the Born-Oppenheimer correction which is significant because of the large mass change of the light Li atom.
    
    We have found a triplet splitting of $p$-wave FRs at high magnetic fields for the $m_N=0$ and $m_N=\pm 1$ components of the scattering channel. By our cc calculation we have attributed this effect to spin-rotation coupling. A simple model to calculate the contributions to this coupling at large $R$ indicates the dominating short-range contributions, highlighting the ability of using FRs to access subtle angular-momentum couplings and molecular wavefunctions in a diatomic molecule. We calculated the binding energies for the hyperfine manifold of the levels directly below the atom pair asymptote $f_{\mathrm{Li}}$=1/2+$f_{\mathrm{Cs}}$=3 at zero magnetic field and found that the spin-rotation contribution for rotational angular momentum up to $l=2$ is below 300~kHz and strongly dependent on the total angular momentum. Measuring these binding energies with high precision, e.g. with rf-spectroscopy, would allow one to determine the sign of this interaction. The extrapolation to higher $N$ values and especially to more deeply-bound levels will lead to significant uncertainties because of missing experimental data in that range. 
    
    \begin{acknowledgements}
    	We are grateful to E. Lippi for helping prepare Fig. \ref{fig:LiCs_sketch} and S. Jochim, F. Ferlaino, S. Whitlock, and P. Fabritius for fruitful discussions. S.H. acknowledges support by the IMPRS-QD. This work is supported in part by the Heidelberg Center for Quantum Dynamics, the DFG/FWF FOR 2247 under Project No. WE2661/11-1, and the DFG Collaborative Research Centre SFB 1225 (ISOQUANT).
    \end{acknowledgements}

\vspace{0.2cm}
B.Z. and S.H. contributed equally to this work.

\section{Appendix}

\subsection{Calculations for the simple model}
\label{appenA}

\subsubsection{Ground state molecular wavefunctions at large $R$}
\label{sec:WF}
    
\paragraph{Electron wavefunctions}
To estimate the WF and energy of the atom pair at large $R$ (see Fig. \ref{fig:LiCs_sketch}) in Eq. (\ref{eq:groundstate}), we follow the perturbative derivation of the dispersion coefficients for the ground-state potentials of bialkali atoms presented in Ref. \cite{Marinescu1995}, which is valid roughly to the LeRoy radius ($R_\mathrm{LR}\approx20$~$a_0$) \cite{LeRoy1974}. 

\begin{figure}[t]
	\centering
	\includegraphics[width=0.5\textwidth]{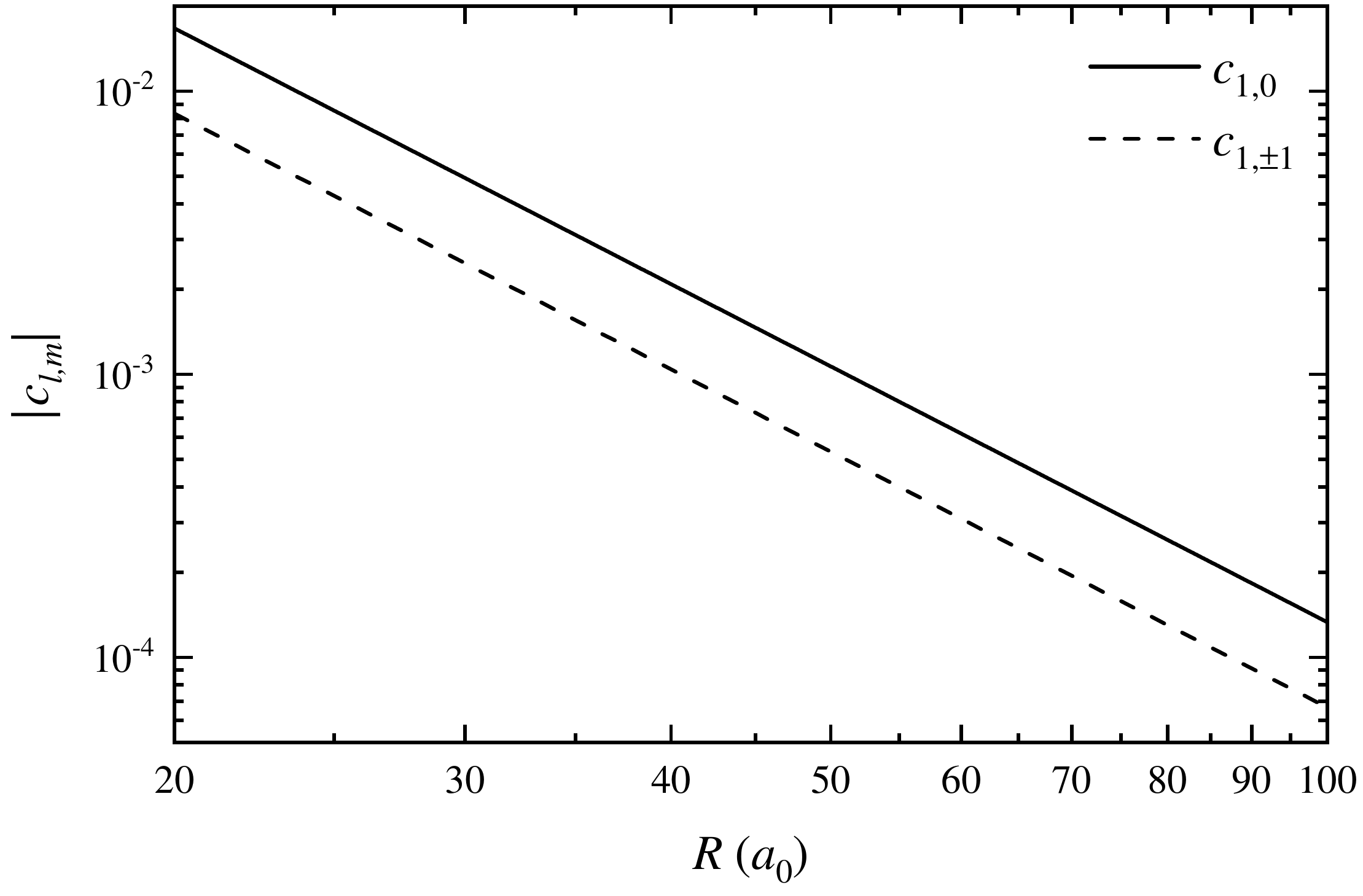}
	\caption{Log-log plot of the absolute values of the mixing coefficients $c_{1,m}$ for the excited state $\ket{\phi^\mathrm{Li}_{2p,m}} \ket{\phi^\mathrm{Cs}_{6p,-m}}$ with $m=0,\pm1$. The slopes of the lines are $-3$ coming from a $R^{-3}$ dependence (see Eq. \eqref{eq:wf_p}).}
	\label{fig:c_coeff}
\end{figure}

The first-order correction of the unperturbed ground state WF $|\Phi_\mathrm{2s,6s}\rangle$ involves the states with both atoms being in their first excited state $|\Phi_\mathrm{2p,6p}\rangle$. More explicitly, the first-order WF reads
\begin{equation}
	\ket{\Phi;R}=\ket{\phi^\mathrm{Li}_{2s}}\ket{\phi^\mathrm{Cs}_{6s}} +\sum\limits_{m=-1,0,1}c_{1,m} \ket{\phi^\mathrm{Li}_{2p,m}} \ket{\phi^\mathrm{Cs}_{6p,-m}} 
	\label{eq:wf_p}
\end{equation}
with $m$ denoting the projection of the electronic orbital momentum. Here, the coefficients are given by $$c_{1,0}=2c_{1,\pm1}\equiv-\frac{2c}{3R^3}(E_{\mathrm{Li,2p}}+E_{\mathrm{Cs,6p}})^{-1}$$ with $c=\braket{\phi^\mathrm{Li}_{2p}|r|\phi^\mathrm{Li}_{2s}} \braket{\phi^\mathrm{Cs}_{6p}|r|\phi^\mathrm{Cs}_{6s}}\approx24.1$, which are plotted in Fig. \ref{fig:c_coeff} as a function of $R$. The second-order contributions are a factor of $10^{-4}$ smaller and thus negligible. The corresponding energy correction is of second-order and represents the well-known van-der-Waals potential $V_\mathrm{vdW} = - C_6/R^6$ for two atoms in the ground state with the coefficient $$C_6=\frac{2c^2}{3( E_{\mathrm{Li,2p}}+E_{\mathrm{Cs,6p}})}\approx3210,$$ which is consistent with the values from our cc calculation (see Table \ref{tab:LiCspot}) and Refs. \cite{Marinescu1995, Marinescu1999}.

\paragraph{Nuclear wavefunction}
For the last vibrational state of the LiCs triplet ground state potential a$^3\Sigma^+$, the inner ($R_i$) and outer ($R_o$) turning points are at about 8 and 43~$a_0$, respectively. Using the semi-classicial (WKB) method \cite{Landau1991}, at $R_i<R<R_o$ the nuclear wavefunction $\psi(R)$ is proportional to $\frac{1}{\sqrt{p(R)}}\sin[\frac{1}{\hbar}\int\limits_{R}^{R_o} p(R)dR + \frac{\pi}{4}]$, while it decays exponentially outside this region. Here $p(R)=\sqrt{2\mu[E_b-V(R)-\frac{N(N+1)}{2\mu R^2}]}$ is the local classical momentum with $E_b\approx-h\times3.4$~GHz the binding energy \cite{Pires2014a}. The molecular potential $V(R)$ dominates over $E_b$ and the centrifugal barrier $\frac{N(N+1)}{2\mu R^2}$ in most cases between the two classical turning points, so $p(R)\approx\sqrt{-2\mu V(R)}$. It can be approximated by the Van-der-Waals potential $V_\mathrm{vdW}=-C_6/R^6$ at $R>R_\mathrm{LR}$, which results in $\psi(R)\propto R^{3/2}$ for $R_\mathrm{LR}<R<R_o$.

\subsubsection{First-order contribution}
As drawn in Fig. \ref{fig:LiCs_sketch}, the magnetic interaction energy of the electron spins with the electronic and nuclear orbital motions is given by \cite{Vleck1951}
    \begin{equation}
    H^{(1)}=\frac{1}{2}g\alpha^2(\sum_{\beta,i}\frac{Z_{i\beta}\mathbf{v}_{\beta}\times\mathbf{r}_{i\beta}}{r_{i\beta}^3}\cdot\mathbf{s}_{i}-\sum_{i\neq j}\frac{\mathbf{v}_{i}\times\mathbf{r}_{ji}}{r_{ji}^3}\cdot\mathbf{s}_{j}),
    \label{eq:first-order}
    \end{equation}
    where $\beta$ represents the atomic cores and $(i,j)$ the (two) active electrons, $Z_{i\beta}$ is the effective nuclear charge of $\beta$ seen by the electron $i$. $\mathbf{r}_{i\alpha}=\mathbf{r}_i-\mathbf{r}_\alpha$ and $\mathbf{r}_{ji}=\mathbf{r}_j-\mathbf{r}_i$. Due to the fast oscillatory nature of the electronic motions, their contributions described by the second term in the bracket of Eq. \eqref{eq:first-order} average to zero. The nuclear velocity $\mathbf{v}_\beta$ can be replaced by $\mathbf{\omega}\times\mathbf{r}_\beta$ or $\frac{\mathbf{N}}{\mu R^2}\times\mathbf{r}_\beta$, where $\mu$ is the reduced mass and $R$ is the internuclear distance. Then Eq. \eqref{eq:first-order}  reads
    \begin{equation}
    \begin{aligned}
    H^{(1)} & =\frac{g\alpha^2}{2\mu R^2}\sum_{\beta,i}\frac{Z_{i\beta}(\mathbf{N}\times\mathbf{r}_\beta)\times\mathbf{r}_{i\beta}}{r_{i\beta}^3}\cdot\mathbf{s}_{i}\\
     & = g\alpha^2B(R)\sum_{\beta,i}Z_{i\beta}\frac{(\mathbf{N}\cdot\mathbf{r}_{i})\mathbf{r}_\beta-(\mathbf{r}_\beta\cdot\mathbf{r}_{i\beta})\mathbf{N}}{r_{i\beta}^3}\cdot\mathbf{s}_{i},
    \end{aligned}
    \label{eq:first-order2}
    \end{equation}
    Here, we use the fact that $\mathbf{N}$ is perpendicular to the internuclear axis in a diatomic molecule, i.e. $\mathbf{N}\cdot\mathbf{r}_\beta=0$. Due to the axial symmetry about the internuclear axis $z$ in the molecule, any terms containing odd powers of $x_i$ and $y_i$ are zero, and Eq. \eqref{eq:first-order2} can be further simplified to 
    \begin{equation}
    \begin{aligned}
    H^{(1)}=-g\alpha^2B(R)\mathbf{N}\cdot\sum_{\beta,i}Z_{i\beta}\frac{z_{\beta}(z_i-z_{\beta})}{r_{i\beta}^3}\mathbf{s}_{i},
    \end{aligned}
    \label{eq:first-order3}
    \end{equation}
    resulting in the dimensionless coupling constant $\gamma^{(1)}$ in Eq. \ref{eq:gamma1}.
    
    The contributions to Eq. \eqref{eq:first-order3} involving the valence electron of Cs have the form of $\gamma_{\text{Cs}}^{(1)} B(R)\mathbf{N}\cdot\mathbf{s}_1$. At large $R$, the molecular WF is approximated by $\ket{\Phi;R}$ and $\gamma_{\text{Cs}}^{(1)}$ reads
    \begin{widetext}
    \begin{equation}
    \begin{aligned}
    \gamma_{\mathrm{Cs}}^{(1)}(R) & = -g\alpha^2 \bra{\Phi;R}[\frac{Z_{1\mathrm{Li}}z_{\mathrm{Li}}(-R+r_{1\mathrm{Cs}}\cos\theta_{\mathrm{Cs}})}{(R^2+r_{1\mathrm{Cs}}^2-2Rr_{1\mathrm{Cs}}\cos\theta_{\mathrm{Cs}})^{3/2}}+\frac{Z_{1\mathrm{Cs}}z_{\mathrm{Cs}}}{r_{1\mathrm{Cs}}^2}\cos\theta_{\mathrm{Cs}}]\ket{\Phi;R} \\
    &
    = \frac{g\alpha^2}{1+\eta} \bra{\Phi;R}[\frac{Z_{1\mathrm{Li}}\eta}{R}\sum\limits_{n=0}^\infty(n+1)P_n(\cos\theta_\mathrm{Cs})(\frac{r_\mathrm{1Cs}}{R})^n+\frac{Z_{1\mathrm{Cs}}R}{r_{1\mathrm{Cs}}^2}\cos\theta_{\mathrm{Cs}}]\ket{\Phi;R} \, ,
    \end{aligned}
    \label{eq:CsFO}
    \end{equation}
    \end{widetext}
    where the coordinates are defined in Fig. \ref{fig:LiCs_sketch}, $Z_{1\mathrm{Li}}=1$ and $Z_{1\mathrm{Cs}}=6.4$, $z_{\text{Li (Cs)}}=\frac{\eta}{1+\eta}R$ ($-\frac{1}{1+\eta}R$) with $\eta=22.1$ the mass ratio, and $P_n(\cos\theta)$ are the Legendre polynomials. Similarly, the contributions involving the Li valence electron have the form of  $\gamma_{\text{Li}}^{(1)} B(R)\mathbf{N}\cdot\mathbf{s}_2$ with 
    \begin{widetext}
    \begin{equation}
    \begin{aligned}
    \gamma_{\mathrm{Li}}^{(1)}(R) = \frac{g\alpha^2}{1+\eta} \bra{\Phi;R}[\frac{Z_{2\mathrm{Cs}}}{R}\sum\limits_{n=0}^\infty(n+1)P_n(-\cos\theta_\mathrm{Li})(\frac{r_\mathrm{2Li}}{R})^n+\frac{Z_{2\mathrm{Li}}R}{r_{2\mathrm{Li}}^2}\cos\theta_{\mathrm{Li}}]\ket{\Phi;R} \, ,
    \end{aligned}
    \label{eq:LiFO}
    \end{equation}
    \end{widetext}
    where $Z_{2\mathrm{Cs}}=1$ and $Z_{2\mathrm{Li}}=1.3$. 
    
    Taking the WF in Eq. \eqref{eq:wf_p}, the angular parts of the integrals in Eqs. \ref{eq:CsFO} and \ref{eq:LiFO} have a general expression of
    \begin{equation}
    \begin{aligned}
    I_{l,m;n} & =\int Y_l^{m*}(\theta, \varphi) P_{n}(\cos\theta) Y_l^m(\theta, \varphi) d\theta d\varphi \\
    & = \frac{2l+1}{2}\int P_l^{-m}(\cos\theta) P_{n}(\cos\theta) P_l^{m}(\cos\theta) d\theta\, .
    \end{aligned}
    \label{eq:angular_int}
    \end{equation}
    Here $Y_l^m(\theta, \varphi) = (-1)^m\sqrt{\frac{2l+1}{4\pi}\frac{(l-m)!}{(l+m)!}}P_l^{m}(\cos\theta)e^{im\varphi}$ are the spherical harmonics with the associated Legendre polynomials $P_l^{m}(\cos\theta)$ and only the $s$ ($l=0,m=0$) and $p$ ($l=1,m=0,\pm1$) states are involved. One can easily obtain $I_{0,0;n}=\delta_{n,0}$, $I_{1,0;n}=\delta_{n,0}+\frac{2}{5}\delta_{n,2}$, and $I_{1,\pm1;n}=1/2(-\delta_{n,0}+\frac{2}{5}\delta_{n,2})$, which leads to $\gamma_{\mathrm{Cs}}^{(1)}(R)=g\alpha^2\frac{\eta}{(1+\eta)R}[1+|c_{1,\pm1}|^2(\frac{7}{2}+\frac{27\braket{r_\mathrm{1Cs}^2}}{5R^2})]$ and $\gamma_{\mathrm{Li}}^{(1)}(R)=g\alpha^2\frac{1}{(1+\eta)R}[1+|c_{1,\pm1}|^2(\frac{7}{2}+\frac{27\braket{r_\mathrm{2Li}^2}}{5R^2})]$. 
    
    Due to the facts that $c_{1,m}\ll1$ and $R\gg r_\mathrm{1Cs},r_\mathrm{2Li}$ at large $R$, the first-order \emph{sr} coupling constant can be approximated as
    \begin{equation}
    \begin{aligned}
    \gamma^{(1)}(R)=[\gamma_{\mathrm{Cs}}^{(1)}(R)+\gamma_{\mathrm{Li}}^{(1)}(R)]/2\approx\frac{g\alpha^2}{2R}\,,
    \end{aligned}
    \label{eq:firstorder_final}
    \end{equation}
    which is plotted as a function of $R$ in Fig. \hyperref[fig:WF_gamma]{5(b)} (orange curve). Weighting the $R$-dependent first-order \emph{sr} coupling by the nuclear WF leads to $\gamma^{(1)}(R)B(R)|R\psi(R)|^2\propto R^2$ for $R>R_\mathrm{LR}$, the large-$R$ part dominating the first-order effect in our simple model.
    
    \subsubsection{Second-order contribution}
    The second-order effect is due to the combined perturbations of the \emph{so} and orbit-rotation coupling. The resulting energy shift is expressed in Eq. \eqref{eq:H2}, which can be rewritten as
    \begin{widetext} 
    \begin{equation}
    H^{(2)}(R)= 2\sum\sum_{i} \frac{\braket{ \Phi;R | A_i\mathbf{l_i}\cdot\mathbf{s}_i |\Phi_\mathrm{2p,6p}}\braket{ \Phi_\mathrm{2p,6p} |B(R)\mathbf{l}_i\cdot\mathbf{N}  | \Phi;R }}{E_\mathrm{Li,2p}+E_\mathrm{Cs,6p}},
    \label{eq:H2_2}
    \end{equation}
    \end{widetext}
    where $\mathbf{l_i}\cdot\mathbf{s}_i=(l_i^+s_i^- + l_i^-s_i^+)/2+l_i^zs_i^z$ and $\mathbf{l_i}\cdot\mathbf{N}=(l_i^+N^- + l_i^-N^+)/2+l_i^zN^z$. To retain the form of $\gamma_i^{(2)}\mathbf{s}_i\cdot\mathbf{N}$, the coupling constant $\gamma_i^{(2)}$ for the electron $i$ reads  
    \begin{equation}
    \begin{aligned}
    \gamma_i^{(2)}(R) & = \frac{1}{2}\sum \frac{\braket{ \Phi;R | A_il_i^+ |\Phi_\mathrm{2p,6p}}\braket{ \Phi_\mathrm{2p,6p} |l_i^-  | \Phi;R }}{E_\mathrm{Li,2p}+E_\mathrm{Cs,6p}}\\
          & =\frac{A_i}{2(E_\mathrm{Li,2p}+E_\mathrm{Cs,6p})}f
    \end{aligned}
    \label{eq:gamma2_2}
    \end{equation}
    with 
    \begin{equation}
    \begin{aligned}
    f & = \sum\braket{ \Phi;R |l_i^+ |\Phi_\mathrm{2p,6p}}\braket{ \Phi_\mathrm{2p,6p} |l_i^-  | \Phi;R }\\
    & =\frac{10c^2}{9R^6(E_\mathrm{Li,2p}+E_\mathrm{Cs,6p})^2}.
    \end{aligned}
    \label{eq:gamma2_3}
    \end{equation}
    Thus $\gamma^{(2)}(R)$ can be estimated as
    \begin{equation}
    \begin{aligned}
    \gamma^{(2)}(R) = \frac{1}{2}\sum_i\gamma_i^{(2)}(R) = \frac{5c^2\sum_iA_i}{18R^6(E_\mathrm{Li,2p}+E_\mathrm{Cs,6p})^3}.
    \end{aligned}
    \label{eq:gamma2_4}
    \end{equation}
    The numerical result of Eq. \eqref{eq:gamma2_4} based on WFs in Appendix \ref{sec:WF} is plotted versus $R$ in Fig. \hyperref[fig:WF_gamma]{5(b)}. Weighting the $R$-dependent second-order \emph{sr} coupling by the nuclear WF leads to $\gamma^{(2)}(R)B(R)|R\psi(R)|^2\propto R^{-3}$ for $R>R_\mathrm{LR}$, the small-$R$ part dominating the second-order effect in our simple model.

 \subsection{Parameters for singlet and triplet ground state potentials of LiCs}
\begin{table*}[h]
		\caption{Potential parameters of the $X^1\Sigma^+$ and $a^3\Sigma^+$ states of LiCs given with respect to the Li(2s)+Cs(6s) asymptote.}
		\label{tab:LiCspot}
		\begin{center}
			\begin{tabular}{|lr|lr|}
				\hline
				\multicolumn{2}{|c|}{$X^{1}\Sigma^{+}$} & \multicolumn{2}{|c|}{$a^{3}\Sigma^{+}$} \\ \hline
				\multicolumn{2}{|c|}{For $R < R_{s} = 2.652 \textnormal{\AA} $} & \multicolumn{2}{|c|}{For $R < R_{s} = 4.345 \textnormal{\AA} $}\\
				A & -0.1013528 $\times 10^{5}$ cm$^{-1}$ &A & -0.36140433 $\times 10^{3}$ cm$^{-1}$ \\
				B & 0.416877623 $\times 10^{6}$ cm$^{-1}$ $\textnormal{\AA}^q$ & B & +  0.847547458 $\times 10^{10}$ cm$^{-1}$ $\textnormal{\AA}^q$\\
				q & 3.81097 & q & 11.55385\\
				\hline
				\multicolumn{2}{|c|}{For $2.652 \textnormal{\AA} = R_{s} \leq R \leq R_{l} = 11.5183 \textnormal{\AA}$} & \multicolumn{2}{|c|}{For $4.345\textnormal{\AA} = R_{s} \leq R \leq R_{l} = 11.5183\textnormal{\AA}$} \\
				b & -0.1 & b & -0.5\\
				$R_m$ & 3.66796875 \AA & $R_m$ & 5.24804688 \AA\\
				$d_0$ & -5875.52991 cm$^{-1}$ & $d_0$ & -309.4623 cm$^{-1}$\\
				$d_1$    &  -0.252618710975387017  $\times 10^{1}$  cm$^{-1}$ & $d_1$    & -0.3642193  cm$^{-1}$ \\                        
				$d_2$    &   0.367457215900665178  $\times 10^{5}$  cm$^{-1}$ & $d_2$    &  0.134782215491288798 $\times 10^{4}$  cm$^{-1}$ \\    
				$d_3$    &   0.435933198105685460  $\times 10^{4}$  cm$^{-1}$ & $d_3$    &  0.537311993871410536 $\times 10^{2}$  cm$^{-1}$\\    
				$d_4$    &  -0.402453029905940202  $\times 10^{5}$  cm$^{-1}$ & $d_4$    & -0.464917969357558093 $\times 10^{3}$  cm$^{-1}$\\    
				$d_5$    &  -0.347796521563863062  $\times 10^{5}$  cm$^{-1}$ & $d_5$    & -0.175749884668625532 $\times 10^{4}$  cm$^{-1}$\\    
				$d_6$    &   0.315821171411499881  $\times 10^{4}$  cm$^{-1}$ & $d_6$    & -0.370441589928719986 $\times 10^{5}$  cm$^{-1}$\\    
				$d_7$    &  -0.266304026591134665  $\times 10^{6}$  cm$^{-1}$ & $d_7$    & -0.278341242922687888 $\times 10^{4}$  cm$^{-1}$\\    
				$d_8$    &  -0.181048129129882972  $\times 10^{7}$  cm$^{-1}$ & $d_8$    &  0.457470834692090517 $\times 10^{6}$  cm$^{-1}$ \\    
				$d_9$    &   0.914704228273577802  $\times 10^{7}$  cm$^{-1}$ & $d_9$    &  0.177854005800842642 $\times 10^{5}$  cm$^{-1}$\\    
				$d_{10}$ &   0.543792503606051728  $\times 10^{8}$  cm$^{-1}$ & $d_{10}$ & -0.276200432485941378 $\times 10^{7}$  cm$^{-1}$\\    
				$d_{11}$ &  -0.259549932529230207  $\times 10^{9}$  cm$^{-1}$ &	$d_{11}$ &  0.431748264218297321 $\times 10^{6}$  cm$^{-1}$\\    
				$d_{12}$ &  -0.105117170881063318  $\times 10^{10}$  cm$^{-1}$ & $d_{12}$ &  0.921774583938866295 $\times 10^{7}$  cm$^{-1}$\\    
				$d_{13}$ &   0.496159149355857754  $\times 10^{10}$  cm$^{-1}$ & $d_{13}$ & -0.348640828981605452 $\times 10^{7}$  cm$^{-1}$\\    
				$d_{14}$ &   0.129151145791594467  $\times 10^{11}$  cm$^{-1}$ & $d_{14}$ & -0.163264106046121176 $\times 10^{8}$  cm$^{-1}$ \\    
				$d_{15}$ &  -0.659976526444605789  $\times 10^{11}$  cm$^{-1}$ & $d_{15}$ &  0.962454228859609924 $\times 10^{7}$  cm$^{-1}$\\    
				$d_{16}$ &  -0.963897189939636536  $\times 10^{11}$  cm$^{-1}$ & $d_{16}$ &  0.120056635752306953 $\times 10^{8}$  cm$^{-1}$\\   
				$d_{17}$ &   0.614868214102462158  $\times 10^{12}$  cm$^{-1}$ & $d_{17}$ & -0.931029427120562643 $\times 10^{7}$  cm$^{-1}$\\   
				$d_{18}$ &   0.341835223965027527  $\times 10^{12}$  cm$^{-1}$ & &\\   
				$d_{19}$ &  -0.393705919840090967  $\times 10^{13}$  cm$^{-1}$ & & \\   
				$d_{20}$ &   0.675695993178769775  $\times 10^{12}$  cm$^{-1}$ & & \\   
				$d_{21}$ &   0.165425882904185527  $\times 10^{14}$  cm$^{-1}$ & &\\   
				$d_{22}$ &  -0.132394231585170703  $\times 10^{14}$  cm$^{-1}$ & &\\   
				$d_{23}$ &  -0.409738756116212656  $\times 10^{14}$  cm$^{-1}$ & &\\   
				$d_{24}$ &   0.607920685011780469  $\times 10^{14}$  cm$^{-1}$ & &\\   
				$d_{25}$ &   0.412071559575104687  $\times 10^{14}$  cm$^{-1}$ & &\\   
				$d_{26}$ &  -0.127770461445448859  $\times 10^{15}$  cm$^{-1}$ & &\\   
				$d_{27}$ &   0.376742627231593516  $\times 10^{14}$  cm$^{-1}$ & &\\   
				$d_{28}$ &   0.957403474032970469  $\times 10^{14}$  cm$^{-1}$ & &\\   
				$d_{29}$ &  -0.940140735588818906  $\times 10^{14}$  cm$^{-1}$ & &\\    
				$d_{30}$ &   0.265036457934536914  $\times 10^{14}$  cm$^{-1}$ & &\\    
				\hline                                                         
				\multicolumn{2}{|c|}{For $R > R_{l} = 11.5183 \textnormal{\textnormal{\AA}} $} &\multicolumn{2}{|c|}{For $R > R_{l} = 11.5183 \textnormal{\textnormal{\AA}} $}\\
				$C_6$ & 0.1486670$\times 10^{8}$ cm$^{-1}$ $\textnormal{\textnormal{\AA}}^ 6$ & $C_6$ & 0.1486670$\times 10^{8}$ cm$^{-1}$ $\textnormal{\textnormal{\AA}}^ 6$\\
				$C_8$ & 0.4343228$\times 10^{9}$ cm$^{-1}$ $\textnormal{\textnormal{\AA}}^8$ & $C_8$ & 0.4343228$\times 10^{9}$ cm$^{-1}$ $\textnormal{\textnormal{\AA}}^8$\\
				$C_{10}$ & 0.1951222$\times 10^{11}$ cm$^{-1}$ $\textnormal{\textnormal{\AA}}^{10}$ & $C_{10}$ & 0.1951222$\times 10^{11}$ cm$^{-1}$ $\textnormal{\textnormal{\AA}}^{10}$\\
				$C_{25}$ & -0.1090612$\times 10^{27}$ cm$^{-1}$ $\textnormal{\textnormal{\AA}}^{25}$ & $C_{25}$ & -0.421260895$\times 10^{25}$ cm$^{-1}$ $\textnormal{\textnormal{\AA}}^{25}$\\
				$A_{ex}$ &0.11330361$\times 10^{6}$ cm$^{-1}$ $\textnormal{\textnormal{\AA}}^{-\gamma}$ & $A_{ex}$ &0.11330361$\times 10^{6}$ cm$^{-1}$ $\textnormal{\textnormal{\AA}}^{-\gamma}$ \\
				$\gamma$ & 5.0568 & $\gamma$ & 5.0568\\
				$\beta$ & 2.2006 $\textnormal{\textnormal{\AA}}^{-1}$ & $\beta$ & 2.2006 $\textnormal{\textnormal{\AA}}^{-1}$\\
				\hline
				\multicolumn{2}{|c|}{For all $R$} & &\\
				$v_0$ & -0.503126 cm$^{-1}$ & &\\
				\hline
			\end{tabular}
		\end{center}
	\end{table*}

\bibliography{Mixtures}
	

\end{document}